\documentstyle[12pt]{article}
\def\asy#1#2{\smash{\mathop{\sim}\limits_{#1}^{#2}}}

\def\ACAL{{\cal A}}

\def\ASF{\hbox{\sf A}}

\def\eqt#1{Eq.(\ref{#1})}
\def\eqts#1{Eqs.(\ref{#1})}
\def\eqtm#1{(\ref{#1})}
\def\vphi{\varphi}

\def\vtheta{\vartheta}
\def\frakd#1#2{{\displaystyle#1\over\displaystyle#2}}

\def\hei{\vphantom{\Bigg|}}

\def\heia{\rule[0pt]{0pt}{5ex}}
\def\heib{\rule[0pt]{0pt}{-5ex}}

\def\cosp#1{\cos^{#1}\hskip-2pt}
\def\sinp#1{\sin^{#1}\hskip-2pt}

\def\Trace#1{{\rm Tr} \hskip-2pt \left( #1 \right)}
\def\Real#1{{\rm Re} \hskip-2pt \left( #1 \right)}
\def\Imag#1{{\rm Im} \hskip-2pt \left( #1 \right)}

\def\MatEl#1#2#3{\left\langle#1\vphantom{#2#3}\right|
	         \vphantom{#1}#2\vphantom{#3}
		 \left|\vphantom{#1#2}#3\right\rangle}

\def\vereq#1#2{\lower3pt\vbox{\baselineskip1.5pt \lineskip1.5pt
               \ialign{$\mpth#1\hfill##\hfil$\crcr#2\crcr\sim\crcr}}}
\def\mpth{\mathsurround=0pt}
\def\CRM{{\rm C}}

\def\epsilonslash{\epsilon\kern-.4em/}
\def\kslash{k\kern-.47em/}
\def\Lslash{L\kern-.45em/}
\def\pslash{p\kern-.435em/}
\def\partialslash{\partial\kern-.53em/}
\def\qslash{q\kern-.46em/}
\def\Rslash{R\kern-.6em/}
\def\sslash{s\kern-.44em/}
\def\vslash{v\kern-.47em/}
\def\xislash{\xi\kern-.44em/}
\def\MeV{\,{\rm M e\hskip-1pt V}}
\def\GeV{\,{\rm G e\hskip-1pt V}}

\def\vet#1{\vec{\hskip+1pt#1}\hskip+1pt}
\def\disty{\displaystyle}

\newcounter{figs}
\newcounter{refs}
\def\mkappa{m}
\begin{document}
\flushbottom
\pagestyle{empty}
\setcounter{page}{0}
\rightline{\sf DFTT 13/93}
\rightline{\sf March 1993}
\rightline{\sf hep-ph/9304221}
\vspace*{1cm}
\begin{center} \LARGE\bf
The Process
$ \bar{p} p \rightarrow e^{-} e^{+} $
with Polarized Initial Particles
and
Proton Form Factors in Time-like Region
\end{center}
\vspace*{1cm}
\centerline{\Large\sf
S.M. Bilenky$^{\rm(a,b,c)}$,
C. Giunti$^{\rm(b,c)\star}$
and
V. Wataghin$^{\rm(b,c)}$ }
\medskip
\centerline{\large
(a) Joint Institute of Nuclear Research, Dubna, Russia }
\medskip
\centerline{\large
(b) INFN Torino, Via P. Giuria 1, I--10125 Torino, Italy }
\medskip
\centerline{\large
(c) Dipartimento di Fisica Teorica, Universit\`a di Torino }
\vspace*{1cm}
\centerline{\Large Abstract }
\bigskip

The process
$ \bar{p} p \rightarrow e^{-} e^{+} $
is considered
in the general case of polarized initial particles.
A relation between the difference of the phases
of the electromagnetic form factors $G_M$ and $G_E$
in the time-like region
and measurable asymmetries
is derived.
It is shown
that the moduli of the form factors
can be determined
from measurements
of the total unpolarized cross section
and of the integral asymmetry
for longitudinally polarized
(or transversely polarized)
$\bar{p}$ and $p$.
The behaviour of the proton form factors
at high $q^2$ in the time-like region
is also discussed.
From the
Phragm\'en-Lindel\"of's theorem
it follows
that the asymptotical behaviour of the form factors
in the space-like and time-like regions
must be the same.
An analysis of experimental data
in both regions
based on perturbative QCD is presented.

\vfill
\noindent
$^{\scriptstyle\star}$
Bitnet: GIUNTI@TORINO.INFN.IT
\newpage
\pagestyle{plain}

The charge and magnetic form factors of the nucleon,
$ G_{E}(q^2) $
and
$ G_{M}(q^2) $,
are classical object of investigation.
For a long time these fundamental quantities
have been investigated in the region of space-like $q^2$.
Starting from the seventies
some informations
on the electromagnetic form factors
of the proton in the time-like region
have been obtained.
Recently rather accurate measurements
of the proton form factors
in the time-like region,
from
$ q^2 = 4 M^2 $
up to
$ q^2 = 4.2 \GeV^2 $,
have been done at LEAR~[\ref{LEAR91a},\ref{LEAR91b}].
Some informations
on the electromagnetic form factor of the proton
at high time-like $ q^2 $
were also obtained
at Fermilab~[\ref{FNAL92}].
Most recent is
the first experimental determination
on the neutron form factor
at $ q^2 \simeq 4.0 \GeV^2 $~[\ref{Adone92}].

There exist several QCD calculations of the electromagnetic form factors
in the space-like region~[\ref{QCD}].
According to our knowledge there are no QCD-based calculations
of the form factors in the time-like region.
The phenomenological models which try to describe
the behaviour of the form factors in both
space-like and time-like regions
are based on the vector meson dominance
model~[\ref{WATAGHIN69}--\ref{BDDS92}].
%[\ref{WATAGHIN69},\ref{CNV82},\ref{BHTZ83},\ref{DUBNICKA},\ref{BDDS92}].
However,
even taking into account all known meson resonances
(and one additional~[\ref{BDDS92}])
it is not possible to obtain a statistically acceptable description
of all the existing experimental data.
Let us remark that the steep decrease
of the form factors
near threshold in the time-like region
discovered in Ref.[\ref{LEAR91a}]
could be explained~[\ref{Dalkarov92}]
by proton-antiproton
interaction in the initial state.

The full understanding of nucleon electromagnetic form factors
still remains a challenge for the theory.
It is clear that any additional information
about the form factors which could be obtained from experiment
is very important.

Taking into account
possible future developments
of the experiments at LEAR~[\ref{LEAP92}]
we analyze here
which additional informations on
the proton form factors in the time-like region
can be obtained from the investigation of the process
\begin{equation}
\bar{p} p \rightarrow e^{-} e^{+}
\label{E1}
\end{equation}
with a polarized proton target
and/or a polarized antiproton beam.
The possibility
to polarize the antiproton beam at LEAR was considered
in Ref.[\ref{PS173}].

Let us consider the process \eqtm{E1}
in the general case
of polarized initial particles.
The matrix element of the process
in the one-photon approximation
has the form
\begin{equation}
\MatEl{ f }{ S }{ i }
=
- i e^2
\overline{u}(k) \gamma_{\alpha} u(-k') \,
\frakd{ 1 }{ q^2 } \,
\overline{u}(-p') \Gamma^{\alpha} u(p)
(2\pi)^4 \delta^4(k+k'-p-p')
\;.
\end{equation}
Here
$p$ and $p'$
are the 4-momenta of the initial proton and antiproton,
$k$ and $k'$
are the 4-momenta of the final electron and positron,
$q=p+p'$
is the 4-momentum of the virtual photon
and
\begin{equation}
\Gamma^{\alpha}
=
\gamma^{\alpha} F_1(q^2)
- \frakd{ i }{ 2 M } \,
\sigma^{\alpha\beta} q_{\beta} F_2(q^2)
\end{equation}
is the electromagnetic vertex of the nucleon
($M$ is the nucleon mass).
The form factors
$F_1(q^2)$
and
$F_2(q^2)$
are connected with the magnetic
$G_M(q^2)$
and charge
$G_E(q^2)$
form factors of the nucleon
by the relations
\begin{equation}
\begin{array}{rcl} \displaystyle \hei
G_M
& = & \displaystyle \hei
F_1 + F_2
\;,
\\ \displaystyle
G_E
& = & \displaystyle
F_1 + \frakd{ q^2 }{ 4 M^2 } \, F_2
\;.
\end{array}
\end{equation}

The cross section of process \eqtm{E1}
in the general case of polarized initial particles
is given by
\begin{equation}
d \sigma
=
\frakd{ 2 \alpha^2 }{ \sqrt{ q^2 ( q^2 - 4 M^2 ) } }
\left( k_{\alpha} k'_{\beta}
- k \cdot k' g_{\alpha\beta}
+ k'_{\alpha} k_{\beta} \right)
\frakd{ 1 }{ q^4 }
\Trace{ \Gamma^{\alpha} \rho(p) \overline\Gamma^{\beta} \rho(-p') }
\delta^4(k+k'-p-p') \,
\frakd{ d^3k }{ k^0 } \,
\frakd{ d^3k' }{ k'^0 }
\;.
\label{E5}
\end{equation}
Here
\begin{equation}
\rho(p)
=
\frakd{1}{2}
\left( 1 - \gamma_5 \xislash \right)
\left( \pslash + M \right)
\end{equation}
is the proton density matrix
($ \xi^{\alpha} $ is the polarization 4-vector of the proton)
and
\begin{equation}
\rho(-p')
=
- \CRM \rho^{T}(p') \CRM^{-1}
\;,
\end{equation}
where
$\CRM$
is the charge-conjugation matrix
and
\begin{equation}
\rho(p')
=
\frakd{1}{2}
\left( 1 - \gamma_5 \xislash' \right)
\left( \pslash' + M \right)
\end{equation}
is the antiproton density matrix
($ \xi'^{\alpha} $ is the polarization 4-vector of the antiproton).

The polarization 4-vector of a particle
in the system where its momentum is $\vet{p}$
is connected with the polarization vector $\vec{P}$
in its rest frame by a Lorentz boost
\begin{equation}
\begin{array}{rcl} \displaystyle \hei
\vec{\xi}
& = & \displaystyle
\vec{P}
+
\frakd{ \vet{p} \left( \vec{P} \cdot \vet{p} \right) }
      { M \left( p^0 + M \right) }
\;,
\\ \displaystyle \hei
\xi^0
& = & \displaystyle
\frakd{ \vec{\xi} \cdot \vet{p} }{ M }
\end{array}
\;.
\label{E9}
\end{equation}

With the help of \eqts{E5}--\eqtm{E9}
we obtain the following expression
for the cross section of the process
$ \bar{p} p \rightarrow e^{-} e^{+} $
in the center of mass system~\footnote
{
The cross section for process \eqtm{E1}
in the case of polarized initial particles
was calculated in Ref.[\ref{Bilenky65}].
We present here the
expression for this cross section in a different form
which is more suitable for our discussion.
Notice that in \eqt{E11} the electron mass has been neglected.
}
\begin{equation}
\begin{array}{rcl} \displaystyle
\left( \frakd{ d \sigma }{ d \Omega } \right)_{\vec{P}';\vec{P}}
& = & \displaystyle
\frakd{ \alpha^2 }{ 4 \sqrt{ q^2 \left( q^2 - 4 M^2 \right) } }
\left\{
\left( 1 + \cosp{2}\vtheta \right) |G_M|^2
+
\sinp{2}\vtheta \, \frakd{ 4 M^2 }{ q^2 } \, |G_E|^2
\right.
\\ \displaystyle
& & \displaystyle
\phantom{ \frakd{ \alpha^2 }{ 4 \sqrt{ q^2 \left( q^2 - 4 M^2 \right) } } \Bigg\{ }
\left.
+
\sin2\vtheta \, \frakd{ 2 M }{ \sqrt{q^2} } \,
\Imag{ G_M G_E^* }
\left[
\left( \vec{P}_{\perp} \cdot \vet{n} \right)
+
\left( \vec{P}'_{\perp} \cdot \vet{n} \right)
\right]
\right.
\\ \displaystyle
& & \displaystyle
\phantom{ \frakd{ \alpha^2 }{ 4 \sqrt{ q^2 \left( q^2 - 4 M^2 \right) } } \Bigg\{ }
\left.
+
\sin2\vtheta \,
\frakd{ 2 M }{ \sqrt{q^2} } \,
\Real{ G_M G_E^* }
\left[
\left( \vec{P}_{\perp} \cdot \vet{s} \right) P'_{\parallel}
-
\left( \vec{P}'_{\perp} \cdot \vet{s} \right) P_{\parallel}
\right]
\right.
\\ \displaystyle
& & \displaystyle
\phantom{ \frakd{ \alpha^2 }{ 4 \sqrt{ q^2 \left( q^2 - 4 M^2 \right) } } \Bigg\{ }
\left.
-
\left[
\left( 1 + \cosp{2}\vtheta \right) |G_M|^2
-
\sinp{2}\vtheta \, \frakd{ 4 M^2 }{ q^2 } \, |G_E|^2
\right]
P_{\parallel} P'_{\parallel}
\right.
\\ \displaystyle
& & \displaystyle
\phantom{ \frakd{ \alpha^2 }{ 4 \sqrt{ q^2 \left( q^2 - 4 M^2 \right) } } \Bigg\{ }
\left.
+
2 |G_M|^2 \sinp{2}\vtheta
\left( \vec{P}_{\perp} \cdot \vet{s} \right)
\left( \vec{P}'_{\perp} \cdot \vet{s} \right)
\right.
\\ \displaystyle
& & \displaystyle
\phantom{ \frakd{ \alpha^2 }{ 4 \sqrt{ q^2 \left( q^2 - 4 M^2 \right) } } \Bigg\{ }
\left.
+
\sinp{2}\vtheta
\left[ \frakd{ 4 M^2 }{ q^2 } \, |G_E|^2 - |G_M|^2 \right]
\vec{P}_{\perp} \cdot \vec{P}'_{\perp}
\right\}
\;.
\end{array}
\label{E11}
\end{equation}
In deriving \eqt{E11}
we expressed
the polarization vectors of the antiproton and proton
(in their rest frame)
as sums of their longitudinal and transverse components
\begin{equation}
\begin{array}{rcl} \displaystyle \hei
\vec{P}'
& = & \displaystyle
\vec{P}'_{\parallel} + \vec{P}'_{\perp}
\;,
\\ \displaystyle \hei
\vec{P}
& = & \displaystyle
\vec{P}_{\parallel} + \vec{P}_{\perp}
\;,
\end{array}
\end{equation}
where
$ \vec{P}'_{\parallel} = P'_{\parallel} \, \vet{\mkappa} $
and
$ \vec{P}_{\parallel} = P_{\parallel} \left( - \vet{\mkappa} \right) $.
Here $\vet{\mkappa}$
is the unit vector in the direction of the momentum
of the antiproton in the c.m.s.
Further,
in \eqt{E11}
$\vtheta$
is the angle between the momenta of the antiproton and the electron,
$ \vet{n} = \vet{\mkappa} \times \vet{k} / | \vet{\mkappa} \times \vet{k} | $
is the unit vector orthogonal to the reaction plane
and
$ \vet{s} = \vet{n} \times \vet{\mkappa} $
is the unit vector in the reaction plane
orthogonal to $\vet{\mkappa}$.

The nucleon form factors in the time-like region
are complex.
In the case of unpolarized initial particles
the cross section depends only on the squared moduli
$|G_M|^2$ and $|G_E|^2$.
As it can be seen from \eqt{E11},
the study of process \eqtm{E1}
with polarized initial particles
could allow to obtain informations
also about the phase difference
$ \chi = \chi_M - \chi_E $,
where
$ \chi_M = {\rm Arg} \, G_M $
and
$ \chi_E = {\rm Arg} \, G_E $,
which
is an important characteristic
of the form factors in the time-like region.

The value of $\sin\chi$
can be obtained from measurements of the cross section
of process \eqtm{E1}
with
an unpolarized antiproton beam
and
a polarized proton target
(or
a polarized antiproton beam
and
an unpolarized proton target).
If the target polarization is orthogonal to the beam direction,
from \eqt{E11} we obtain the asymmetry
\begin{equation}
\begin{array}{rcl} \displaystyle \hei
\ACAL(\vtheta,\vphi)
& = & \displaystyle
\left[
\frakd{
\left( \frakd{ d \sigma }{ d \Omega } \right)_{0;\vec{P}_{\perp}}
-
\left( \frakd{ d \sigma }{ d \Omega } \right)_{0;-\vec{P}_{\perp}}
}{
\left( \frakd{ d \sigma }{ d \Omega } \right)_{0;\vec{P}_{\perp}}
+
\left( \frakd{ d \sigma }{ d \Omega } \right)_{0;-\vec{P}_{\perp}}
}
\right]
\frakd{ 1 }{ |\vec{P}_{\perp}| }
\\ \displaystyle \hei
& = & \displaystyle
\frakd{ \heia
\sin2\vtheta \, \cos\vphi \,
\frakd{ 2 M }{ \sqrt{q^2} } \,
|G_M| |G_E| \sin\chi
}{
\left( 1 + \cosp{2}\vtheta \right)
|G_M|^2
+
\frakd{ 4 M^2 }{ q^2 } \,
\sinp{2}\vtheta
|G_E|^2
}
\;,
\end{array}
\label{E23}
\end{equation}
where $\vphi$ is the angle between
$ \vec{P}_{\perp} $ and $ \vet{n} $.
Let us notice that
the asymmetry $\ACAL(\vtheta,\vphi)$ must vanish
at the threshold
$q^2=4M^2$,
where
$ G_M = G_E $.
Notice also that in the case of unpolarized proton target
and polarized antiproton beam
the asymmetry is the same
as in the case of polarized target and unpolarized beam.
This is a consequence of invariance
under charge conjugation.

Information about $\cos\chi$
can be obtained from measurements
of the cross section of process \eqtm{E1}
with
a transversely polarized beam
and a longitudinally polarized target
(or
a longitudinally polarized beam
and a transversely polarized target).
If the antiproton polarization vector is
$ \vec{P}'_{\perp} $
and the proton polarization vector
is
$ - P_{\parallel} \vet{\mkappa} $
($ P_{\parallel} \vet{\mkappa} $),
for the asymmetry we have
\begin{equation}
\begin{array}{rcl} \displaystyle \hei
\ACAL_{\perp;\parallel}(\vtheta,\vphi')
& = & \displaystyle
\left[
\frakd{
\left( \frakd{ d \sigma }{ d \Omega }
\right)_{\vec{P}'_{\perp};-P_{\parallel}\vet{\mkappa}}
-
\left( \frakd{ d \sigma }{ d \Omega }
\right)_{\vec{P}'_{\perp};P_{\parallel}\vet{\mkappa}}
}{
\left( \frakd{ d \sigma }{ d \Omega }
\right)_{\vec{P}'_{\perp};-P_{\parallel}\vet{\mkappa}}
+
\left( \frakd{ d \sigma }{ d \Omega }
\right)_{\vec{P}'_{\perp};P_{\parallel}\vet{\mkappa}}
}
\right]
\frakd{ 1 }{ |\vec{P}'_{\perp}| \, P_{\parallel} }
\\ \displaystyle \hei
& = & \displaystyle
-
\frakd{ \heia
\sin2\vtheta \, \sin\vphi' \,
\frakd{ 2 M }{ \sqrt{q^2} } \,
|G_M| |G_E| \cos\chi
}{
\left( 1 + \cosp{2}\vtheta \right)
|G_M|^2
+
\frakd{ 4 M^2 }{ q^2 } \,
\sinp{2}\vtheta
|G_E|^2
}
\;,
\end{array}
\label{E24}
\end{equation}
where $\vphi'$ is the angle between
$ \vec{P}'_{\perp} $ and $ \vet{n} $.

It is obvious that the differential asymmetry $ \ACAL(\vtheta,\vphi) $
is maximal if the target polarization vector
$ \vec{P}_{\perp} $
is directed
along $ \vet{n} $ (i.e. $ \vphi = 0 $).
Analogously the asymmetry $ \ACAL_{\perp;\parallel}(\vtheta,\vphi') $
is maximal if the beam polarization vector
$ \vec{P}'_{\perp} $
is directed along
$ \vet{s} $
(i.e. $ \vphi' = \pi/2 $).
In this case,
from \eqts{E23} and \eqtm{E24}
we have
\begin{equation}
\tan\chi
=
-
\frakd{ \ACAL(\vtheta,0) }{ \ACAL_{\perp;\parallel}(\vtheta,\pi/2) }
\;,
\label{E25}
\end{equation}
which
is a relation
between the phase difference $ \chi $
and measurable asymmetries.

Up to now we have considered differential asymmetries.
It is not difficult to generalize
\eqt{E25} for integral asymmetries.
It is
obvious that the terms
$ \left( \vec{P}_{\perp} \cdot \vet{n} \right) $
and
$ \left( \vec{P}'_{\perp} \cdot \vet{s} \right) P_{\parallel} $
in expression \eqtm{E11} for the cross section
vanish after integration
over the total solid angle.
However,
we can consider the asymmetries
integrated over part of the solid angle.
Let us consider the process \eqtm{E1}
with an unpolarized beam and
a target with polarization
$ \vec{P}_{\perp} $
orthogonal to the momentum of the antiproton
and integrate the cross section over the angle $\vphi$
between
$ \vec{P}_{\perp} $
and
$\vet{n}$
from
$ - \pi/2 $ to $ \pi/2 $
and over the angle $\vtheta$
from $0$ to $\pi/2$.
From \eqt{E11}, for the integral asymmetry we have
\begin{equation}
\ASF
=
\frakd{
\frakd{ 4 M }{ \pi \sqrt{q^2} } \,
|G_M| |G_E| \sin\chi
}{
2 |G_M|^2
+
\frakd{ 4 M^2 }{ q^2 } \,
|G_E|^2
}
\;.
\label{E26}
\end{equation}

Now let us consider process \eqtm{E1}
with a longitudinally polarized proton target
with polarization
$ - P_{\parallel} \vet{\mkappa} $
($ P_{\parallel} \vet{\mkappa} $),
and a transversely polarized antiproton beam
with polarization $ \vec{P}'_{\perp} $
and integrate the differential cross section
over the angle $\vphi'$ between $ \vec{P}'_{\perp} $
and $\vet{n}$
from $0$ to $\pi$
and over the angle $\vtheta$
from $0$ to $\pi/2$.
For the asymmetry we have
\begin{equation}
\ASF_{\perp;\parallel}
=
-
\frakd{
\frakd{ 4 M }{ \pi \sqrt{q^2} } \,
|G_M| |G_E| \cos\chi
}{
2 |G_M|^2
+
\frakd{ 4 M^2 }{ q^2 } \,
|G_E|^2
}
\;.
\label{E27}
\end{equation}

From \eqts{E26} and \eqtm{E27}
we obtain the following relation between
the phase difference $\chi$
and the integral asymmetries
\begin{equation}
\tan\chi
=
-
\frakd{ \ASF }{ \ASF_{\perp;\parallel} }
\;.
\label{E28}
\end{equation}
Thus measurements of the integral asymmetries
$ \ASF $
and
$ \ASF_{\perp;\parallel} $
would allow us to determine the phase difference $\chi$
directly from experimental data.
Notice that
the phase difference $\chi$
can be determined unambiguously
with this method
(in addition to \eqt{E28}
it is necessary to take into account the sign of
$ \ASF $
or
$ \ASF_{\perp;\parallel} $).
Let us stress that the knowledge
of the moduli
$|G_M|$ and $|G_E|$
is not necessary
in order to obtain $\chi$
with the help of \eqt{E28}.

As it can be seen from \eqt{E11}
if both the initial particles
have longitudinal or transverse polarizations,
then the integral cross section depends only on the moduli
of the form factors.
Let us consider first the case of longitudinal polarizations.
From \eqt{E11}
for the asymmetry
integrated over the total solid angle
we have
\begin{equation}
\begin{array}{rcl} \displaystyle \heib
\ASF_{\parallel;\parallel}
& = & \displaystyle
\left[
\frakd{
\sigma_{ P'_{\parallel} \vet{\mkappa} ; - P_{\parallel} \vet{\mkappa} }
-
\sigma_{ P'_{\parallel} \vet{\mkappa} ; P_{\parallel} \vet{\mkappa} }
}{
\sigma_{ P'_{\parallel} \vet{\mkappa} ; - P_{\parallel} \vet{\mkappa} }
+
\sigma_{ P'_{\parallel} \vet{\mkappa} ; P_{\parallel} \vet{\mkappa} }
}
\right]
\frakd{ 1 }{ P'_{\parallel} \, P_{\parallel} }
\\ \displaystyle \heia
& = & \displaystyle
-
\frakd{
2 |G_M|^2 - \frakd{ 4 M^2 }{ q^2 } \, |G_E|^2
}{
2 |G_M|^2 + \frakd{ 4 M^2 }{ q^2 } \, |G_E|^2
}
\;.
\end{array}
\label{E14}
\end{equation}
This asymmetry depends only on $|G_M|^2$ and $|G_E|^2$.
The value of $|G_M|^2$ and $|G_E|^2$
can be determined from the measurement of the
$\cosp{2}\vtheta$ dependence
of the differential cross section of process \eqtm{E1}
in the case of unpolarized initial particles.
However,
this method requires
a rather high statistical accuracy of the data.
Here we discuss an alternative method
for the determination of $|G_M|^2$ and $|G_E|^2$.
If the asymmetry $ \ASF_{\parallel;\parallel} $
is measured,
these quantities can be determined directly
from the value of $ \ASF_{\parallel;\parallel} $ 
and the value of the total cross section
of process \eqtm{E1}
with unpolarized initial particles
\begin{equation}
\sigma_0
=
\overline\sigma
\left[
2 |G_M|^2 + \frakd{ 4 M^2 }{ q^2 } \, |G_E|^2
\right]
\;,
\label{E15}
\end{equation}
where
\begin{equation}
\overline\sigma
=
\frakd{ 2 \pi \alpha^2 }{ 3 \sqrt{ q^2 \left( q^2 - 4 M^2 \right) } }
\;.
\end{equation}
Indeed from \eqts{E14} and \eqtm{E15} we have
\begin{equation}
\begin{array}{rcl} \displaystyle \hei
|G_M|^2
& = & \displaystyle
\frakd{ 1 }{ 4 }
\left( 1 - \ASF_{\parallel;\parallel} \right)
\frakd{ \sigma_0 }{ \overline\sigma }
\;,
\\ \displaystyle \heia
|G_E|^2
& = & \displaystyle
\frakd{ q^2 }{ 8 M^2 }
\left( 1 + \ASF_{\parallel;\parallel} \right)
\frakd{ \sigma_0 }{ \overline\sigma }
\;.
\end{array}
\end{equation}

As can be seen from \eqt{E11},
the cross section of process \eqtm{E1}
depends only on $|G_M|^2$ and $|G_E|^2$
also
if both beam and target have transverse polarizations.
In this case,
assuming that the polarizations
of the proton and antiproton
are parallel (antiparallel),
for the integral asymmetry we have
\begin{equation}
\ASF_{\perp;\perp}
=
\frakd{
\frakd{ 4 M^2 }{ q^2 } \, |G_E|^2
}{
2 |G_M|^2 + \frakd{ 4 M^2 }{ q^2 } \, |G_E|^2
}
\;.
\label{E17}
\end{equation}
With the help of \eqts{E15} and \eqtm{E17}
we obtain
\begin{equation}
\begin{array}{rcl} \displaystyle \hei
|G_M|^2
& = & \displaystyle
\frakd{ 1 }{ 2 }
\left( 1 - \ASF_{\perp;\perp} \right)
\frakd{ \sigma_0 }{ \overline\sigma }
\;,
\\ \displaystyle \hei \heia
|G_E|^2
& = & \displaystyle
\frakd{ q^2 }{ 4 M^2 } \,
\ASF_{\perp;\perp} \,
\frakd{ \sigma_0 }{ \overline\sigma }
\;.
\end{array}
\end{equation}
So an investigation of the process
$ \bar{p} p \rightarrow e^{-} e^{+} $
in the case of longitudinally polarized
or transversely polarized
beam and target
will allow us to determine
the moduli of the charge and magnetic proton form factors
from measurements of the {\bf integral} cross section.

The experiments discussed here
without any doubt
are very difficult.
However,
taking into account the importance of the informations
which could be obtained
from the investigation of polarization effects
in process \eqtm{E1},
we think that it is appropriate to discuss
the possibility of their measurement~\footnote
{
Let us notice that the proton polarization
in the process
$ e^{-} e^{+} \rightarrow \bar{p} p $
was discussed in Ref.[\ref{Dubnicka92}].
}.

In conclusion
we would like to make the following remarks

\begin{enumerate}

\item
As it can be seen from \eqt{E11},
in the cross section of process \eqtm{E1}
with unpolarized particles
$ \disty \left( \frakd{ d \sigma }{ d \Omega } \right)_{0;0} $
the value of $|G_E|^2$
is multiplied by
$ 4 M^2 / q^2 $.
This means that at high $q^2$
the main contribution to
$ \disty \left( \frakd{ d \sigma }{ d \Omega } \right)_{0;0} $
comes from $|G_M|^2$.
On the other hand,
from \eqt{E27}
it follows that at high $q^2$
\begin{equation}
\ASF_{\perp;\parallel}
\simeq
-
\frakd{ 2 M }{ \pi \sqrt{q^2} } \,
\frakd{ |G_E| }{ |G_M| } \,
\cos\chi
\;.
\label{E18}
\end{equation}
So measurements of the asymmetry
$ \ASF_{\perp;\parallel} $
allow to determine the ratio
$ |G_E| / |G_M| $
at high $q^2$
(if the phase $\chi$ is known from measurements
of the asymmetries
$\ASF$
and
$ \ASF_{\perp;\parallel} $).

\item
In order to illustrate the possible behaviour
of the considered asymmetries,
we calculate them using
the parametrizations of the form factors
proposed in
Ref.[\ref{BHTZ83}]
and
Ref.[\ref{BDDS92}].
These parametrizations are based
on the vector meson dominance model.
The parameters were determined in
Ref.[\ref{BHTZ83}]
and
Ref.[\ref{BDDS92}]
from a fit of existing experimental data.
The results of our calculations
of the asymmetries
$ \ASF $
and
$ \ASF_{\perp;\parallel} $
are presented in Fig.\ref{FIG1} and Fig.\ref{FIG2},
respectively.
It can be seen from Fig.\ref{FIG1} and Fig.\ref{FIG2}
that different models predict quite different
behaviours of the asymmetries.

\item
Since
$ G_E(4M^2) = G_M(4M^2) $,
it is clear that the asymmetry
$ \ASF $
vanishes at the threshold.
It is possible to show that
the asymmetry
$ \ASF $
goes to zero at
$ q^2 \to \infty $.
In fact the electromagnetic form factors
$ G_{E,M}(q^2) $
are
limiting values of the functions
$ G_{E,M}(z) $
\begin{equation}
G_{E,M}(q^2)
=
\lim_{\epsilon\to0^{+}} G_{E,M}(q^2+i\epsilon)
\label{E31}
\end{equation}
which are analytical in the upper half
of the complex $z$ plane
and increase at infinity not faster than a power of $z$.
We can apply~[\ref{Logunov65}]
to the form factors
the Phragm\'en-Lindel\"of's theorem~[\ref{Titchmarsh}]
which was used~[\ref{Sugawara61},\ref{Meiman63}]
to proof the Pomeranchuk theorem
and its generalizations~[\ref{Logunov65}].
From this theorem it follows that
the form factors have the same asymptotical behaviour
in the space-like and time-like regions.
In the space-like region
the form factors are real.
This means that
the form factors are real in the time-like region
at asymptotically high $ q^2 $
and
from \eqt{E23}
it follows that
$ \ASF \to 0 $
at $ q^2 \to \infty $.

\item
Let us discuss in some more detail
the asymptotical behaviour
of the electromagnetic form factors of the nucleon
in the time-like region.
In accordance with the quark counting rule
[\ref{Matveev73},\ref{BrodskyFarrar73}],
which is based on the scaling hypothesis and dimensional arguments,
at high $ |q^2| $
the form factors of the nucleon
(system of three quarks)
behave as
\begin{equation}
G_{M}(q^2)
\sim
F_{1}(q^2)
\sim
\frakd{1}{q^4}
\;.
\label{E32}
\end{equation}
The quark-gluon interaction
leads to violation of scaling
and additional logarithmic $q^2$ dependence of the form factors.
Let us write
in the space-like region ($q^2<0$)
\begin{equation}
\frakd{ G_{M}(q^2) }{ \mu_p }
\ \asy{q^2\to-\infty}{} \ 
\frakd{ C_s }{ q^4 } \, \Phi(q^2)
\;,
\label{E33}
\end{equation}
where $\mu_p=2.79$
is the proton magnetic moment
in nuclear magnetons.
From
the leading order perturbative QCD
it follows that
$ \Phi(q^2) $
is given by~[\ref{BL}]
\begin{equation}
\Phi(q^2)
=
\alpha_s^2(-q^2)
\left[ \ln\left(
- \frakd{ q^2 }{ \Lambda^2 }
\right) \right]^{-4/3\beta}
\;,
\label{E34}
\end{equation}
where
\begin{equation}
\alpha_s(-q^2)
=
\frakd{ 4 \pi }{ \beta \ln\left( - \frakd{q^2}{\Lambda^2} \right) }
\;,
\qquad\qquad
\beta = 11 - {2\over3} \, n_f
\;,
\label{E35}
\end{equation}
$n_f$
is the number of flavours
and
$\Lambda$
is the QCD scale parameter.
The value of the constant
$C_s$
is determined by the wave function of the
nucleon~[\ref{BL}--\ref{GS}].
%[\ref{BL},\ref{CZ},\ref{GS}].
The Phragm\'en-Lindel\"of's theorem implies
that
in the time-like region ($q^2>0$)
the form factors have the following asymptotical behaviour
\begin{equation}
\frakd{ G_{M}(q^2) }{ \mu_p }
\ \asy{q^2\to\infty}{} \ 
\frakd{ C_t }{ q^4 } \, \Phi(q^2)
\label{E36}
\end{equation}
with
\begin{equation}
C_t = C_s
\;.
\label{E37}
\end{equation}

Let us compare
the behaviour of the form factors
at large space-like and time-like
momentum transfer.
In a recent SLAC experiment~[\ref{SLAC92}]
the elastic electron-proton cross section
was measured in a wide range of momentum transfer,
from
$ -q^2 = 2.9 \GeV^2 $
to
$ -q^2 = 31.3 \GeV^2 $.
From these measurements
the values of the form factor $G_{M}(q^2)$
at high $-q^2$
can be extracted
(the contribution to the cross section
of the form factor $G_{E}(q^2)$
at high $-q^2$
is small and can be neglected).
In Fig.\ref{FIG3}
the experimental points for
$ |G_{M}| q^4 / \mu_p $
at $-q^2>10\GeV^2$
are plotted together with
curves obtained from fits
of the data using \eqt{E33}.
With
$ \Lambda = 100 \MeV $
we obtain
$ C_s = 12.3 \pm 0.2 \GeV^4 $
($ \chi^2 / {\rm N D F} = 4.6 / 5 $)
and with
$ \Lambda = 200 \MeV $
we get
$ C_s = 7.8 \pm 0.1 \GeV^4 $
($ \chi^2 / {\rm N D F} = 10.2 / 5 $).
Let us notice that the quality of the fit
depends rather strongly
on the value of $\Lambda$
(smaller values of
$ \Lambda $
are preferable).

The cross section of the process
$ \bar{p} p \rightarrow e^{-} e^{+} $
at high $ q^2 $
($ q^2 = 8.9 $, $ 12.4 $, $ 13.0 \GeV^2 $)
was measured recently
in a Fermilab experiment~[\ref{FNAL92}].
In Fig.\ref{FIG3} we have plotted
the values of
$ |G_{M}| q^4 / \mu_p $
obtained from this experiment.
There are also shown
two fits of these data
made using \eqt{E36}
with
$ \Lambda = 100 \MeV $
($ C_t = 30.9^{+4.1}_{-4.8} \GeV^4 $,
 $ \chi^2 / {\rm N D F} = 0.29 / 1 $)
and
$ \Lambda = 200 \MeV $
($ C_t = 21.2^{+2.8}_{-3.3} \GeV^4 $,
 $ \chi^2 / {\rm N D F} = 0.29 / 1 $).
We have taken into account~\footnote
{
We are grateful to Dr. S. Brodsky
for calling our attention
to the possible numerical importance
of the imaginary part of the form factor
in the range of values of $q^2$
presently available.
Notice that at $q^2\simeq10\GeV^2$
the relative contribution to $|G_M|$
of the imaginary part of $G_M$
is about 20\%.
}
the imaginary part of the form factor
in the time like region
that arises due to
$ \disty
\ln\left( - \frakd{ q^2 }{ \Lambda^2 } \right)
=
\ln\left( \frakd{ q^2 }{ \Lambda^2 } \right)
+ i \pi
$.
Thus the experimental data in the space-like
as well as in the time-like regions of $q^2$
are described by expressions \eqtm{E33} and \eqtm{E36},
respectively.
The accuracy of the data
in the time-like region
is much worse
than that
in the space-like region.
As a consequence,
the corresponding
accuracy of the determination
of the constant $C_t$
in the time-like region
is much worse
than that
of the constant $C_s$
in the space-like region.
However,
the average values of $C_t$ and $C_s$
are so different that,
even with such a low accuracy in the determination of $C_t$,
we can conclude
that these constants are different
($C_t$ is more than $3\sigma$ higher than $C_s$).
From our point of view
this difference means that
the range of high $q^2$ values
investigated in present experiments is not asymptotic
and the Phragm\'en-Lindel\"of's theorem
does not apply in this range.
Nonperturbative effects~[\ref{IL}],
nonleading log corrections
and other effects
could be important
in this region.

Further investigations
of the processes
$ e p \to e p $
and
$ \bar{p} p \rightarrow e^{-} e^{+} $
and comparisons of the behaviour
of the form factors
in the regions of space-like and time-like
high $q^2$
seems to us very important
from the point of view of the possibility
to test the perturbative QCD prediction
of the asymptotical behaviour
of the proton form factor
in a way which does not
dependent on the choice
of the nucleon wave function.

\end{enumerate}

\bigskip
It is a pleasure for us to thank prof. V. de Alfaro
for fruitful discussions.
We also would like to thank S. Bilenkaya for help in the treatment of
experimental data.
%
%\newpage
\vskip1in
{\Large\bf References }

\begin{list}{[\therefs]}{\usecounter{refs}}

\item\label{LEAR91a}
G. Bardin et al.,
Phys. Lett. B 255 (1991) 149.

\item\label{LEAR91b}
G. Bardin et al.,
Phys. Lett. B 257 (1991) 514.

\item\label{FNAL92}
T. Armstrong et al.,
Phys. Rev. Lett. 70 (1993) 1212.

\item\label{Adone92}
A. Antonelli et al., June 1992.

\item\label{QCD}
For a recent review see:
P. Kroll,
Invited talk given at the IV$^{\rm th}$
International Symposium on Pion-Nucleon Physics
and the Structure of the Nucleon,
Bad Honnef,
September 1991.

\item\label{WATAGHIN69}
V. Wataghin,
Nucl. Phys. B 10 (1969) 107.

\item\label{CNV82}
P. Ceselli, M. Nigro and C. Voci,
Workshop on Physics at LEAR with Low-Energy Cooled Antiprotons,
Erice 1982,
p.~365.

\item\label{BHTZ83}
V. Bardek, Z. Hlousek, W.P. Trower and N. Zovko,
Il Nuovo Cimento A 75 (1983) 368.

\item\label{DUBNICKA}
S. Dubni\v{c}ka,
Il Nuovo Cimento A 100 (1988) 1, 103 (1990) 469;
Il Nuovo Cimento A 103 (1990) 1417, 104 (1991) 1075.

\item\label{BDDS92}
S.I. Bilenkaya, S. Dubni\v{c}ka,
A.Z. Dubni\v{c}kov\'{a} and P. Str\'{\i}\v{z}enec,
Il Nuovo Cimento A 105 (1992) 1421.

\item\label{Dalkarov92}
O.D. Dalkarov and K.V. Protasov,
Phys. Lett. B 280 (1992) 117;
see also Ref.[\ref{LEAR91a}].

\item\label{LEAP92}
See Proceedings of the Second Biennial Conference on Low-Energy
Antiproton Physics LEAP '92,
Courmayeur, Italy,
September 1992.

\item\label{PS173}
The PS 173 Collaboration,
Proceedings of the $3^{\rm th}$ LEAR Workshop,
Tignes 1985.

\item\label{Bilenky65}
S.M. Bilenky and R.M. Ryndin,
Sov. J. Nucl. Phys. 1 (1965) 57.

\item\label{Dubnicka92}
S. Dubni\v{c}ka,
A.Z. Dubni\v{c}kov\'{a}, P. Str\'{\i}\v{z}enec
and M.P. Rekalo,
Proceedings of Hadron Structure '92,
Star\'{a} Lesn\'{a},
Czecho-Slovakia,
September 1992.

\item\label{Logunov65}
A.A. Logunov, N. van Hieu and I.T. Todorov,
Annals of Physics 31 (1965) 203.

\item\label{Titchmarsh}
E.C. Titchmarsh,
``The theory of functions'',
Oxford University Press,
London 1939.

\item\label{Sugawara61}
M. Sugawara and A. Kanazawa,
Phys. Rev. 123 (1961) 1895.

\item\label{Meiman63}
N.N. Me\v{\i}man,
Sov. Phys. JETP 16 (1963) 1609.

\item\label{Matveev73}
V.A. Matveev, R.M. Muradyan and A.N. Tavkhelidze,
Lettere al Nuovo Cimento 7 (1973) 719.

\item\label{BrodskyFarrar73}
S.J. Brodsky and G.R. Farrar,
Phys. Rev. Lett. 31 (1973) 1153.

\item\label{BL}
S.J. Brodsky and G.P. Lepage,
Phys. Rev. Lett. 43 (1979) 545, 1625;
Phys. Rev. D 22 (1980) 2157;
Physica Scripta 23 (1981) 945.

\item\label{CZ}
V.L. Chernyak and I.R. Zhitnitsky,
Nucl. Phys. B246 (1984) 52.

\item\label{GS}
M.Gari and N.G. Stefanis,
Phys. Lett. B 175 (1986) 462;
Phys. Rev. D 35 (1987) 1074.

\item\label{SLAC92}
R.G. Arnold et al.,
Phys. Rev. Lett. 57 (1986) 174;
A.F. Sill et al.,
SLAC-PUB-4395, October 1992.

\item\label{IL}
N. Isgur and C.H. Llewellyn Smith,
Phys. Rev. Lett 52 (1984) 1080;
Phys. Lett. B 217 (1989) 535.

\end{list}
%
%\newpage
\vskip1in
{\Large\bf Figure Captions}

\begin{list}{Fig.\thefigs}{\usecounter{figs}}

\item\label{FIG1}
Integral asymmetry $\ASF$
for the case of unpolarized antiproton beam
and polarized proton target
as a function of $q^2$
in different models:
BHTZ (Ref.[\ref{BHTZ83}]),
BDDS21 (Ref.[\ref{BDDS92}], Table 2, Column 1),
BDDS31 (Ref.[\ref{BDDS92}], Table 3, Column 1),
BDDS32 (Ref.[\ref{BDDS92}], Table 3, Column 2).

\item\label{FIG2}
Integral asymmetry $ \ASF_{\perp;\parallel} $
for the case of transversely polarized antiproton beam
and longitudinally polarized proton target
as a function of $q^2$
in different models:
BHTZ (Ref.[\ref{BHTZ83}]),
BDDS21 (Ref.[\ref{BDDS92}], Table 2, Column 1),
BDDS31 (Ref.[\ref{BDDS92}], Table 3, Column 1),
BDDS32 (Ref.[\ref{BDDS92}], Table 3, Column 2).

\item\label{FIG3}
$ |G_{M}| q^4 / \mu_p $
for
$|q^2|>10\GeV^2$
in the space-like and time-like regions.
The experimental points
are taken from Ref.[\ref{SLAC92}] and Ref.[\ref{FNAL92}].
The curves 
are obtained from fits
of the data using \eqt{E33}
with
$ \Lambda = 100 \MeV $
and
$ \Lambda = 200 \MeV $.

\end{list}

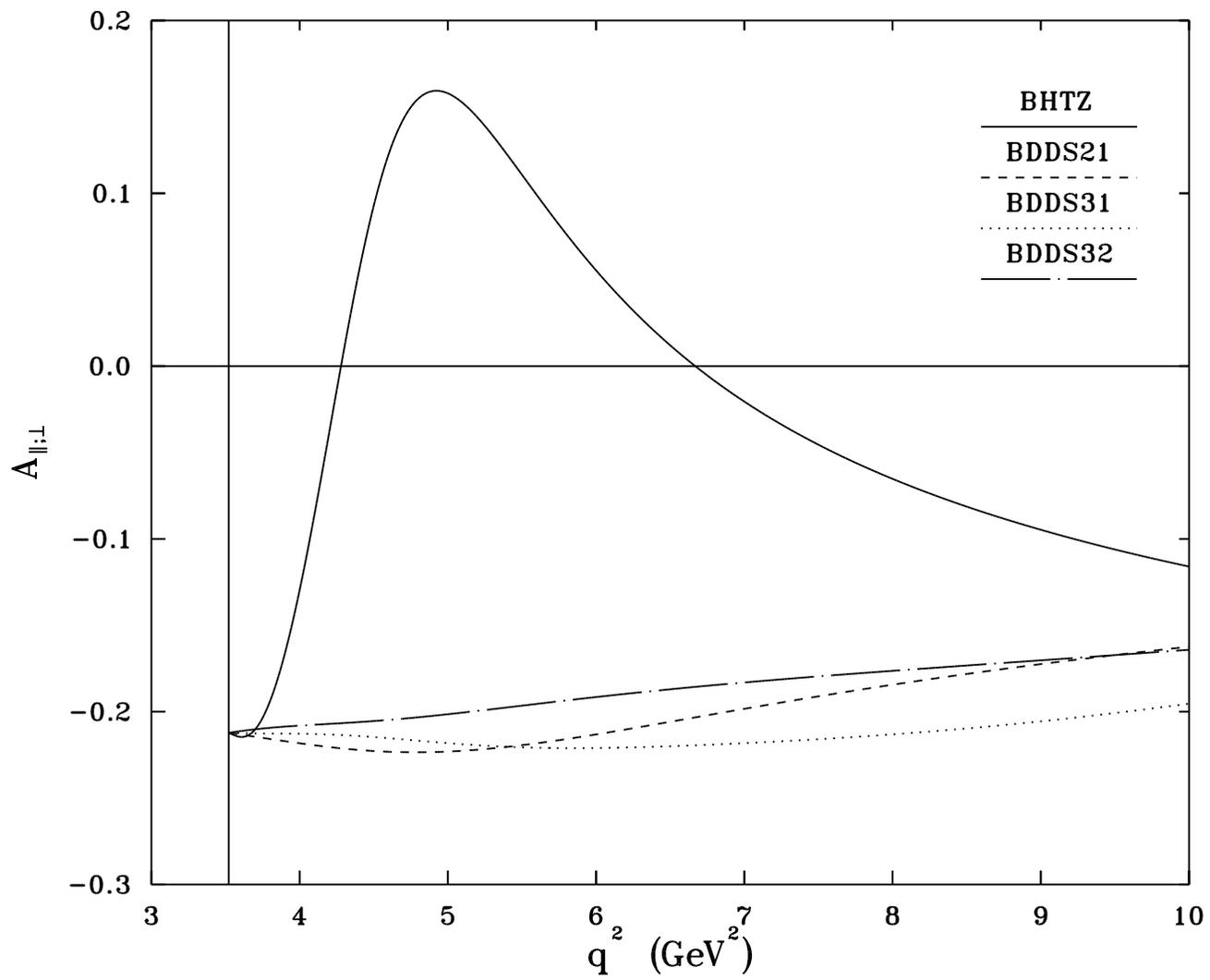
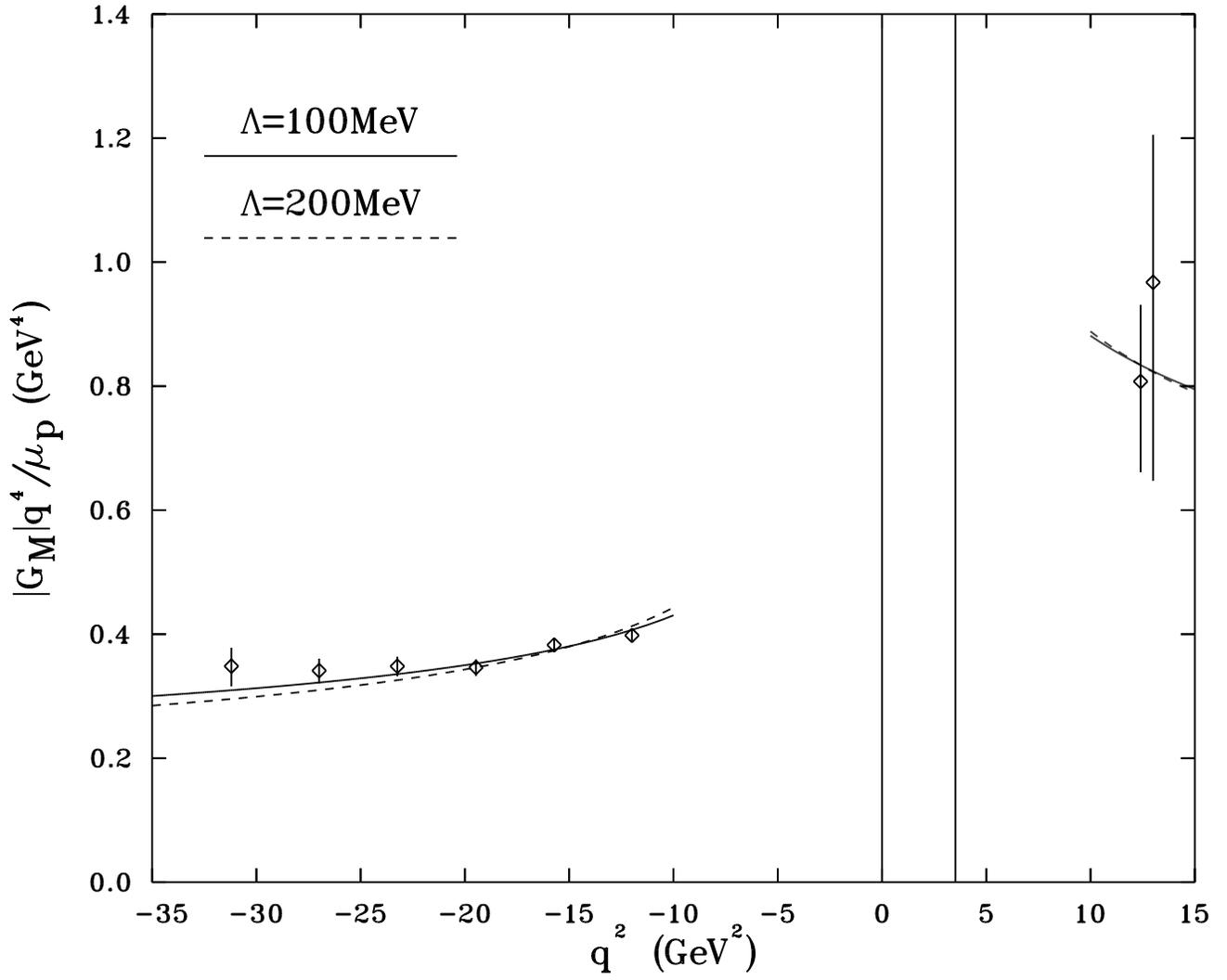
\end{document}